\documentclass[10pt,twocolumn,letterpaper]{article}

\usepackage{cvpr}              

\usepackage{times}
\usepackage{epsfig}
\usepackage{graphicx}
\usepackage{amsmath}
\usepackage{amssymb}
\usepackage{booktabs}
\usepackage{cuted}
\usepackage{capt-of}
\usepackage{multirow}
\usepackage{stfloats}
\usepackage{algorithm2e}
\usepackage{mathtools}
\usepackage{makecell}
\usepackage[accsupp]{axessibility}

\usepackage[dvipsnames]{xcolor}

\usepackage{pifont}
\newcommand{\cmark}{\ding{51}}%
\newcommand{\xmark}{\ding{55}}%

\definecolor{Acolor}{HTML}{16A085}
\definecolor{Bcolor}{HTML}{E67E22}
\definecolor{Ccolor}{HTML}{9B59B6}
\definecolor{Dcolor}{HTML}{2980B9}
\definecolor{Ecolor}{HTML}{3E3E3E}
\definecolor{Fcolor}{HTML}{F1C40F}
\definecolor{Gcolor}{HTML}{C0392B}

\newcommand{\vIfmtZLV}{\textbf{\textcolor{Acolor}{ A.1~}}}
\newcommand{\JnHtnpTW}{\textbf{\textcolor{Acolor}{ A.2~}}}
\newcommand{\vIfmtZLVStageTwo}{\textbf{\textcolor{Acolor}{ A.3~}}}
\newcommand{\JnHtnpTWStageTwo}{\textbf{\textcolor{Acolor}{ A.4~}}}

\newcommand{\ErTWgScg}{\textbf{\textcolor{Bcolor}{ B.1~}}}
\newcommand{\fHNHoNvT}{\textbf{\textcolor{Bcolor}{ B.2~}}}
\newcommand{\sWdhUbZf}{\textbf{\textcolor{Bcolor}{ B.3~}}}
\newcommand{\uapnFuFg}{\textbf{\textcolor{Bcolor}{ B.4~}}}
\newcommand{\MZZCMvBJ}{\textbf{\textcolor{Bcolor}{ B.5~}}}

\newcommand{\PiShi}{\textbf{\textcolor{Ccolor}{ C.1~}}}
\newcommand{\KAraFs}{\textbf{\textcolor{Ccolor}{ C.2~}}}

\definecolor{cvprblue}{rgb}{0.21,0.49,0.74}
\usepackage[pagebackref,breaklinks,colorlinks,citecolor=cvprblue]{hyperref}

\def\paperID{8782} 
\def\confName{CVPR}
\def\confYear{2024}

\title{Looking Similar, Sounding Different: Leveraging Counterfactual Cross-Modal Pairs for Audiovisual Representation Learning}

\author{Nikhil Singh$^{1}$\thanks{Most of the work conducted during author's internship at Netflix.} \qquad Chih-Wei Wu$^{2}$ \qquad Iroro Orife$^{2}$ \quad Mahdi Kalayeh$^{2}$ \vspace{3pt} \\ 
	$^{1}$Massachusetts Institute of Technology \qquad $^{2}$Netflix\\
	$^{1}${\tt\small nsingh1@mit.edu} \quad $^{2}${\tt\small [chihweiw, iorife, mkalayeh]@netflix.com}
}

\begin{document}
\maketitle

\begin{strip}
    \centering
    \vspace{-4em}
    \includegraphics[width=\linewidth]{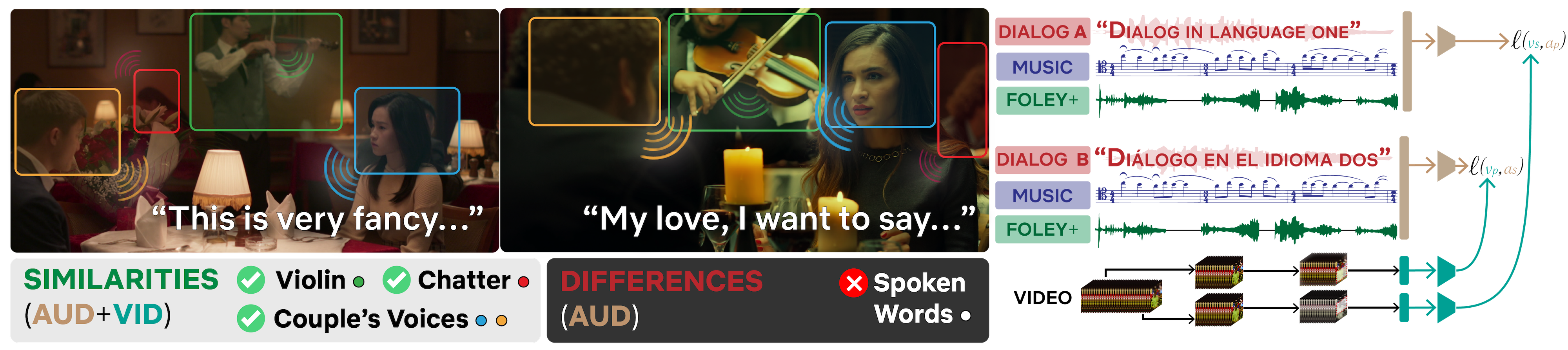}
    \captionof{figure}{
    \textbf{(Left)} Audiovisual scenes can be perceptually similar even as the words spoken in them differ, which may be a challenge for self-supervised audiovisual representation learning. \textbf{(Right)} We propose to leverage movie dubs during training and show that it improves the quality of learned representations on a wide range of tasks.
    }
    \label{fig:banner}
\end{strip}

\setcounter{tocdepth}{-1}

\begin{abstract}
Audiovisual representation learning typically relies on the correspondence between sight and sound. However, there are often multiple audio tracks that can correspond with a visual scene. Consider, for example, different conversations on the same crowded street. The effect of such counterfactual pairs on audiovisual representation learning has not been previously explored. To investigate this, we use dubbed versions of movies and television shows to augment cross-modal contrastive learning. Our approach learns to represent alternate audio tracks, differing only in speech, similarly to the same video. Our results, from a comprehensive set of experiments investigating different training strategies, show this general approach improves performance on a range of downstream auditory and audiovisual tasks, without majorly affecting linguistic task performance overall. These findings highlight the importance of considering speech variation when learning scene-level audiovisual correspondences and suggest that dubbed audio can be a useful augmentation technique for training audiovisual models toward more robust performance on diverse downstream tasks.
\end{abstract}    
\section{Introduction}
\label{sec:intro}
\textit{Can two videos look similar while sounding different?} Consider the two scenes on the left in Fig.~\ref{fig:banner}. These come from different sources, but share elements like a violinist in the background, other tables further away, and a couple's voices in an upscale restaurant environment; but what are they saying? This can vary considerably between the two scenes, even without changing other aspects. General-purpose self-supervised audiovisual representations are often focused on non-speech applications, evidenced by both existing training datasets and common downstream evaluation tasks. In audio alone, there is a myriad of applications beyond semantic speech processing, leading to recent benchmarks which evaluate generalization across and trade-offs between types of tasks~\cite{wang2022towards,turian2022hear}. How then can we focus on learning robust representations from audiovisual content with speech mixed into it? Importantly, there are many non-semantic, or \textit{paralinguistic}, speech processing tasks of interest, as speech is much more than audible text. These too require discovering other similarities beyond words.

\begin{figure}
    \centering
    \includegraphics[width=\linewidth]{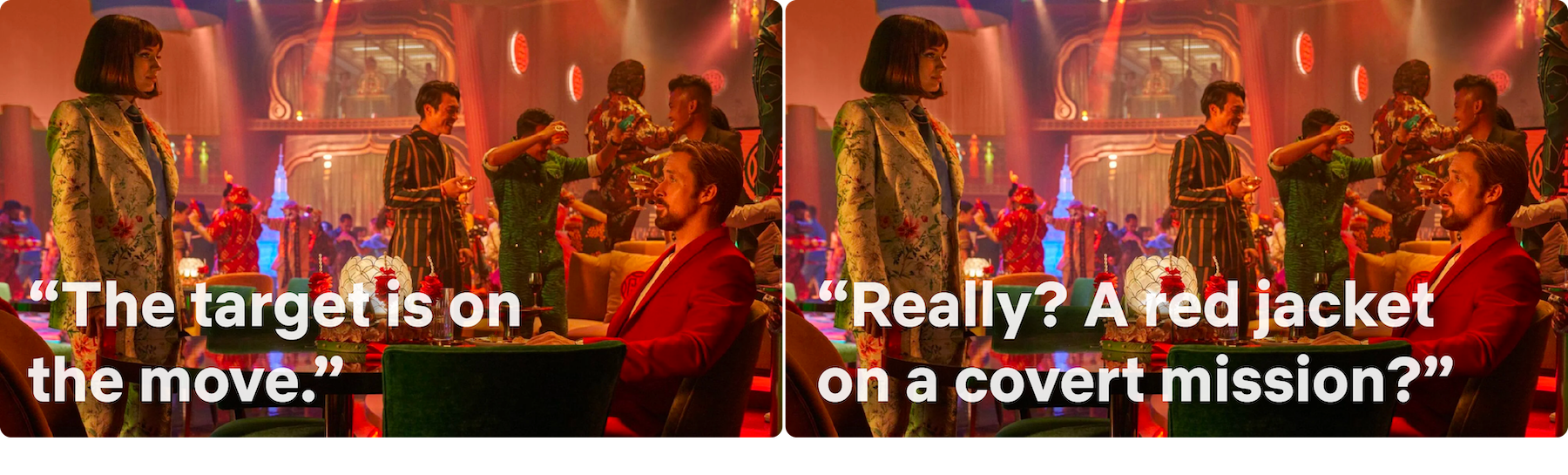}
    \caption{Consider the pictured scene. Which of these dialog examples is more likely? Both are plausible within the scene, yet their phonetic-acoustic characteristics would create differences in the soundtrack.
    }
    \label{fig:example}
\end{figure}

Imagine a movie discussion scene, as in Fig.~\ref{fig:example}. Many audiovisual elements are present: background chatter, glasses clinking, music, footsteps, and characters' voices, but \textit{a priori} this scene could contain many different dialogs without changing the fundamental scene attributes, beyond local features such as lip movements, and this indicates an explicitly counterfactual structure. Note that there are also other counterfactual cross-modal structures which relate to different problems, such as multiple videos of dancing to the same music. Differences in spoken words are one specific case of this which we explore.

In this work, we hypothesize that this \textit{looking similar, while sounding different} problem, as it can occur in real-world audiovisual data distributions, may inhibit the performance of self-supervised audiovisual representation learners. Established approaches, such as cross-modal contrastive learning, where models learn to discriminate true audiovisual pairs from false ones, could be affected; linguistically different but otherwise similar audio-video pairs could act as confounders in this case. However, counterfactual versions of exactly the same scene with only different dialog are generally not available, even if the distribution of real-world audiovisual scenes exhibits this overall trend.

We propose to leverage a data source which naturally resembles this counterfactual-like structure as a proxy: \textit{dubs}. Dubs are alternate versions of movie audio tracks where the speech is replaced with a second-language adaptation, and the rest of the sounds are generally unchanged. Recent works have shown how training on movie scenes can yield strong performance~\cite{kalayeh2022negative, chen2023movies2scenes}, since they contain  diverse audiovisual mixtures, compared with popular audiovisual datasets which are curated to focus on specific objects or actions. Although this distribution may help in learning representations focused on overall scene attributes rather than the dialog's semantics, which is our goal, contrastive training on aligned audio and video from movies does not explicitly account for scenes that look similar and sound different due to linguistic variation. We improve upon this strategy by leveraging multilingual dubbed versions of movies\footnote{The pretraining data also includes episodes of television shows. To avoid clutter, we refer to all long-form content as movies unless it is necessary to specify.}. Specifically, we create a dataset of movies and television shows, each with up to seven audio tracks: English (\textbf{EN}), Spanish (\textbf{ES}), French (\textbf{FR}), Japanese (\textbf{JA}), German (\textbf{DE}), Italian (\textbf{IT}) and Korean (\textbf{KO}). We plug our training strategy into a well-established self-supervised contrastive learning formulation, \textit{i.e.}~SimCLR~\cite{chen2020simple}, and we show that this can improve performance in both multimodal and unimodal setups. Overall, this work contributes:

\begin{itemize}
    \item An approach to improving self-supervised audiovisual representation learning using \textit{dubs}, secondary audio language versions of movies.
    \item Extensive experiments showing that this approach not only improves performance on a range of auditory and audiovisual tasks but also yields new state-of-the-art on multiple benchmarks.
    \item Additional experiments to investigate potential trade-offs. These show that we can get an improvement without majorly affecting the performance on language identification, and semantic speech tasks.
    \item An example pipeline for producing counterfactual pairs in various languages; we apply the workflow to the LVU~\cite{lvu2021} dataset and demonstrate the possibility of creating alternate audio tracks that potentially empower the research community to further investigate the impact of spoken words in audiovisual representation learning.
\end{itemize}

\section{Related Work}
\label{sec:relatedwork}
\paragraph{Self-supervised and Multimodal Learning}
Self-supervised learning relies on pretext tasks with engineered supervision based on data structure, rather than human labels, to learn useful representations~\cite{zhang2016colorful,gidaris2018unsupervised,noroozi2016unsupervised,doersch2015unsupervised,misra2020self,bachman2019learning,hadsell2006dimensionality,he2020momentum}. We focus on \textit{contrastive learning}, which has shown strong performance by maximizing mutual information between views of the same instance~\cite{chen2020simple,he2020momentum,gao2021simcse,tian2020contrastive,bachman2019learning,tschannen2019mutual,tian2020contrastive}. These can then be adapted to novel tasks by fine-tuning, or by appending simple (often linear) models, both with smaller-scale task-specific supervision requirements. \textit{Cross-modal} contrastive learning specifically leverages multimodal data like image and text~\cite{radford2021learning}, or, as in our case, video and audio~\cite{nagrani2021attention,wang2021multimodal,fayek2020large,yang2020telling,alwassel2019self,owens2016ambient,korbar2018cooperative,owens2018audio,arandjelovic2017look,aytar2016soundnet,hu2019deep,nagrani2020speech2action,xiao2020audiovisual,patrick2021compositions,ma2020active}.

\paragraph{Audiovisual Learning}
\vspace{-1em}
Audiovisual learning harnesses cross-modal correspondences for tasks like action~\cite{korbar2018cooperative,kazakos2019epic} and speaker~\cite{nagrani2020voxceleb,chung2018voxceleb2} recognition,  source separation~\cite{rivet2014audiovisual,chatterjee2021visual,tzinis2020into}, media synthesis~\cite{guo2021ad,gan2020foley,owens2016visually,su2020audeo}, audio spatialization~\cite{morgado2018self,garg2021geometry,xu2021visually}, acoustic simulation~\cite{singh2021image2reverb,chen2022visual,majumder2022few}, and more. Much work takes a contrastive approach, recognizing that audio and video can be treated as two complementary sensory views of a single underlying phenomenon, and focuses on learning \textit{coordinated}~\cite{baltruvsaitis2018multimodal} representations. Prior work has found that cross-modal training can lead to better results than within-modal training~\cite{morgado2021audio}, so we use this cross-modal setup as the basis for our framework. In this work, we rely on multilingual audio dubs and videos from long-form content, e.g. movies and television shows. Movies contain rich audiovisual correspondences mimicking real-world experiences, and are more diverse and novel than user-generated videos while being abundant and scalable~\cite{chen2023movies2scenes,tapaswi2016movieqa,huang2020movienet}.

\paragraph{General-purpose Audio Representation Learning and Evaluation}
\vspace{-1em}
Sound is heterogeneous, with speech, music, and environmental sounds having very different characteristics. Even within speech, for example, tasks like speech recognition~\cite{liu2020mockingjay,chi2021audio} and speech emotion recognition~\cite{shneiderman2000limits} differ dramatically. This has motivated developing general-purpose audio representations~\cite{saeed2021contrastive,niizumi2021byol} and benchmarks like HARES~\cite{wang2022towards} and HEAR~\cite{turian2022hear}. We focus our audio evaluation on HEAR~\cite{turian2022hear} since it provides a consistent API. The central hypothesis is that if dub-augmented training in the cross-modal setting improves the generality of the representations, performance on various tasks should increase while avoiding a significant trade-off on language-related tasks.

\paragraph{Multilingual Audio}
\vspace{-1em}
Multilingual speech processing has enabled progress in areas like speech recognition~\cite{byrne2000towards} through pretraining on diverse data~\cite{ghoshal2013multilingual,conneau2020unsupervised}. Recently, speech-to-speech translation has been possible as well~\cite{lee2021textless}. Speech translation in audiovisual media is often referred to as \textit{dubbing}. This is a type of audiovisual translation~\cite{chaume2020audiovisual} in which speech content from a media artifact (\textit{e.g.}~a movie) is re-recorded in another language. Dubs predominate over subtitles in many cultures~\cite{chaume2013turn}. This provides naturalistic multilingual data at scale, and offers a specific case for our hypothesis about audio-visual consistency: a dub's soundtrack differs from the original only in spoken language. We seek to leverage dubs' parallel primary and secondary audio, differing only in speech, to learn more robust audiovisual representations. We also produce a synthetic pipeline for creating counterfactual pairs, to demonstrate the concept of counterfactual cross-modal pairs, while enabling future exploration and validation from the research community.

\section{Pretraining Dataset}
\label{sec:pretraining_dataset}

\begin{figure}
     \centering
     \includegraphics[width=\columnwidth]{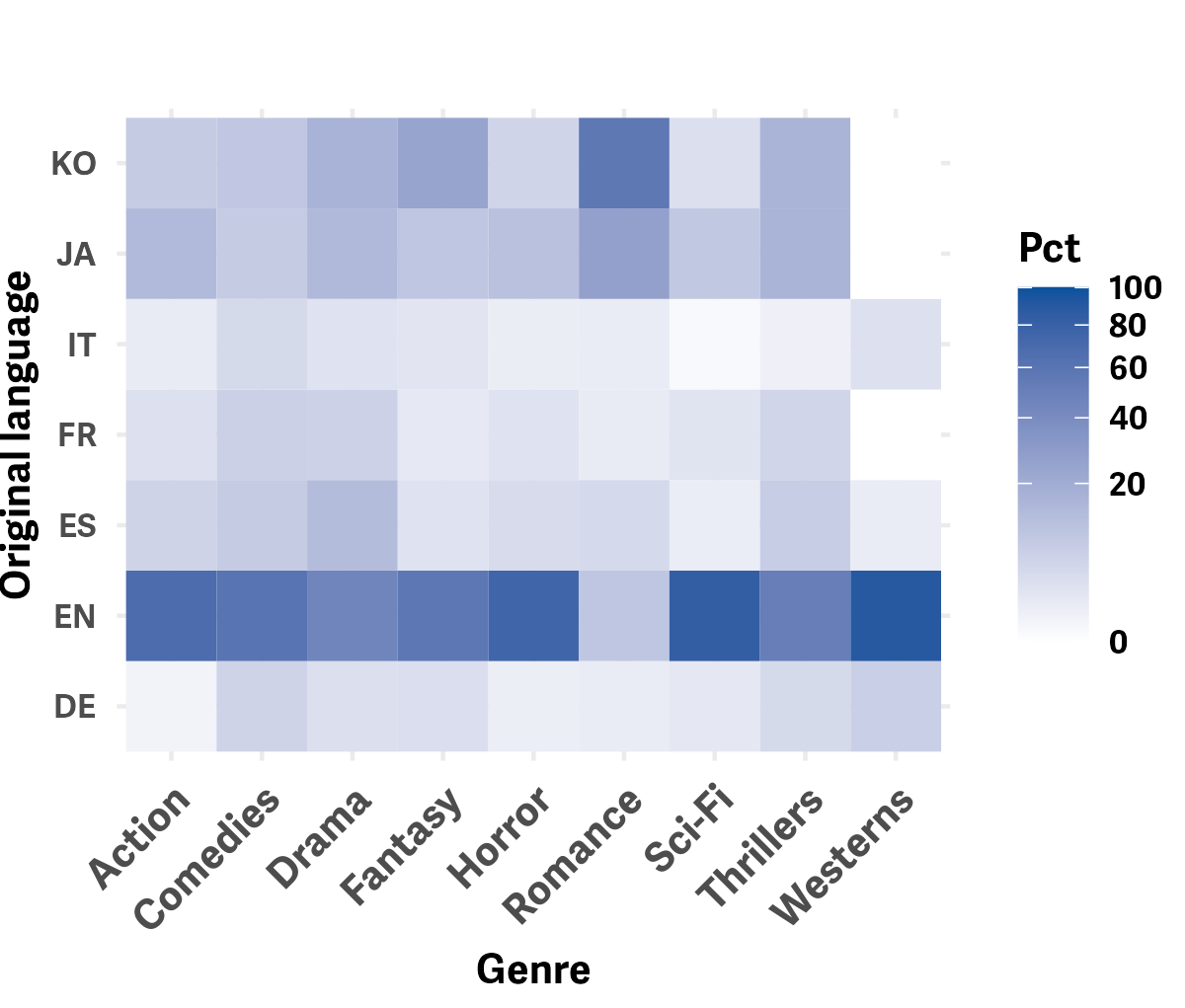}
     \caption{Movies and television episodes included in our pretraining dataset are chosen from a diverse set of original languages and genres. Our goal is to minimize potential content and story biases that could potentially impact our self-supervised models. Note that beyond curating the dataset, we do not use this metadata for representation learning. We normalize per column for visualization.
     }
    \label{fig:dataset_stat}
\end{figure}

\begin{figure*}[!htb]
    \centering
    \includegraphics[width=\textwidth]{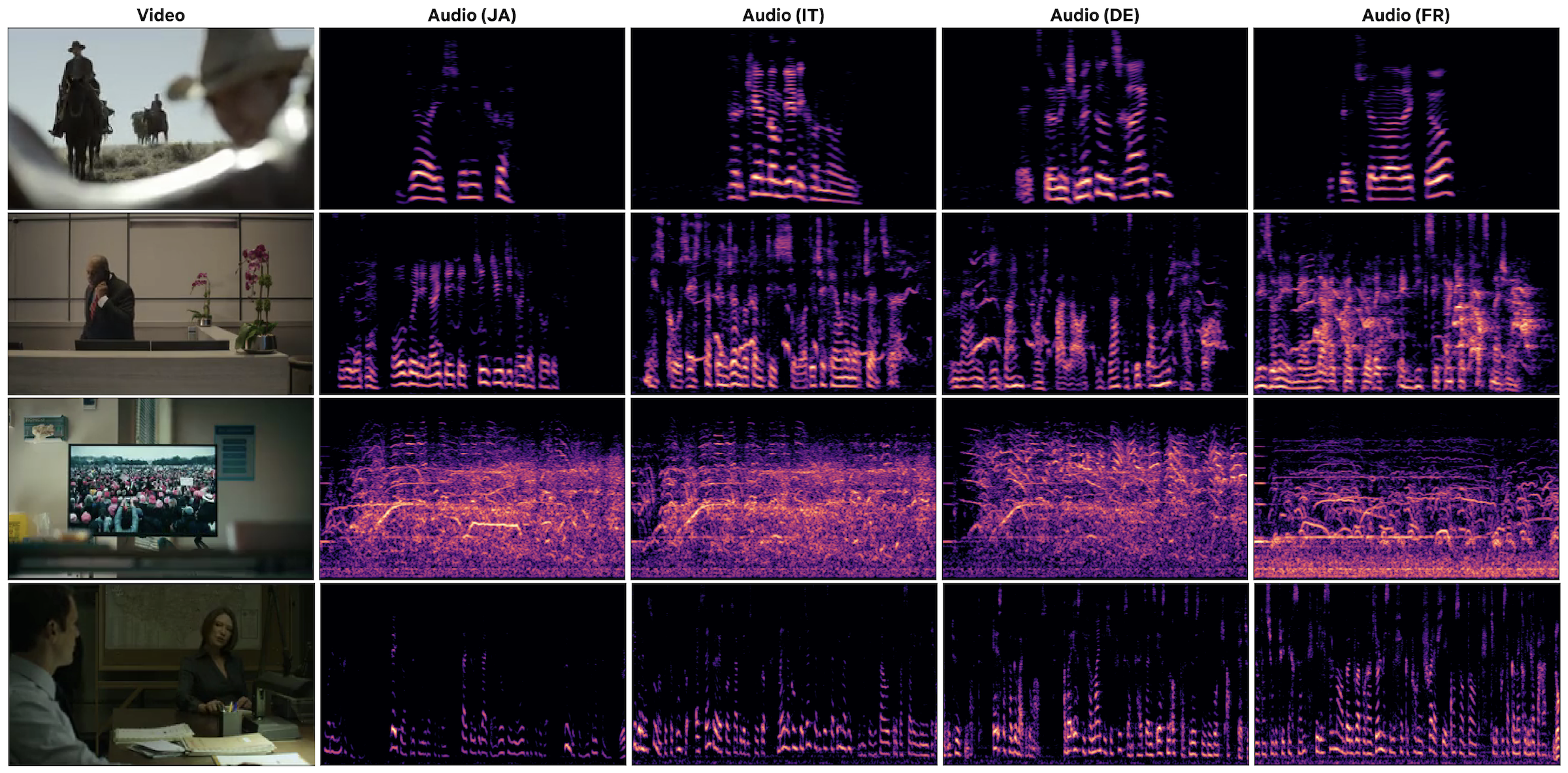}
    \caption{Example clips from our pretraining dataset, showing video stills and mel spectrograms for each of the audio tracks.
    }
    \label{fig:clips}
\end{figure*}

Our dataset consists of $\sim$20K movies and $\sim$33K television episodes, which constitutes $\sim$59K video-hours in total. We have paid extra attention to the diversity of titles used in our pretraining dataset in order to minimize the potential implicit biases in our learned representations, and limited ourselves to only a small part of the catalog to investigate this question. Fig.~\ref{fig:dataset_stat} provides details on the distribution of genre, and original language of the titles included in our dataset\footnote{Further details are given in the appendix.}. Each title contains a video track, as well as up to seven audio tracks: English (\textbf{EN}), Spanish (\textbf{ES}), French (\textbf{FR}), Japanese (\textbf{JA}), German (\textbf{DE}), Italian (\textbf{IT}) and Korean (\textbf{KO}). Most titles have only a single audio track, which is almost always their original language while about a quarter of the dataset is multilingual where on average 2.8 audio tracks are available for each title. Such a dataset allows us to explore the impact of spoken words in audio for self-supervised audiovisual representation learning. Having multiple dub options enables us to investigate trade-offs between secondary languages, and whether ``multilingual'' models might further strengthen downstream performance. 

We recognize that this kind of data has the potential to significantly benefit research. We are actively investigating the necessary legal steps to potentially release a variant of it for non-commercial use. Fig.~\ref{fig:clips} illustrates a few samples from our dataset but readers are encouraged to check out our project page for more examples\footnote{\href{https://nikhilsinghmus.github.io/lssd/}{nikhilsinghmus.github.io/lssd}}.
\section{Methodology}
\label{sec:methodology}

\subsection{Approach} 
Our pretraining dataset is denoted by $\mathcal{X}=\{\mathcal{X}_{n}| n\in[1\cdots N]\}$, where $\mathcal{X}_{n}=\{x_{n,m}| m\in[1\cdots M_{n}]\}$ contains $M_{n}$ non-overlapping snippets which are temporally segmented from the duration of the $n^{th}$ title in the dataset. $\mathcal{Q}$ is a function class which we use to create quadruplet training instances $(v_p, a_p, v_s, a_s)\sim\mathcal{Q}(x_{n,m})$\footnote{subscripts stand for primary and secondary} where $v_p$ and $v_s$ are obtained through spatio-temporal augmentation of video modality in $x_{n,m}$. Similarly are $a_p$ and $a_s$ for the audio modality, yet, unlike video, we do have the opportunity to further add dub-augmentation to audio instances. When more than one language is available this would ensure that $a_p$ and $a_s$ are similar except in their spoken language.

Randomly sampling negatives, the traditional approach in metric and contrastive learning, has been observed to be suboptimal~\cite{saunshi2019theoretical,ma2020active}. A number of recent works develop methods for so-called \textit{hard negative mining}, where the goal is to populate the negative set with challenging examples~\cite{oh2016deep,robinson2020contrastive}. In our case, the data is hierarchical; snippets are naturally nested within source long-form titles, and those from the same title share several common attributes including characters, places, objects, voices, and aesthetics. Hence, following prior work~\cite{kalayeh2022negative}, to create a mini-batch $\mathcal{B}=\{x_i|i\in[1\cdots B]\}$, we first uniformly sample a title, $n\sim\mathbb{U}(1,N)$, and then draw multiple distinct snippets from $\mathcal{X}_{n}$. This ensures that for each instance in $\mathcal{B}$, there are always a sufficient number of samples from the same title to act as hard negatives. This is important since $B\ll N$, hence for $n\sim\mathbb{U}(1,N)$ and $m\ne m'$, $\mathsf{P}(x_{n,m}\in \mathcal{B}\land x_{n,m'}\in \mathcal{B})\to0$. In other words, the naive random sampling policy of $x_i\sim\bigcup_{n=1}^{N}\mathcal{X}_{n}$ would mainly lead to easy cross-title negatives.

We can now formulate the training objective. Considering a cross-modal setup, $\mathcal{B}=\{(v_i,a_i)|i\in[1\cdots B]\}$ represents a minibatch of size $B$, where video and audio modalities of the $i^{th}$ instance are denoted by $v_i$ and $a_i$. We use $z^{i}_{v}$ and $z^{i}_{a}$ to represent their respective embeddings. For the $i^{th}$ element in the minibatch, $(z^{i}_{v},z^{i}_{a})$ serves as the positive pair, while assuming negative pairs for both modalities, $\mathcal{N}_i=\{(z^{i}_{v},z^{j}_{a}),(z^{j}_{v},z^{i}_{a})|j\in[1\cdots B],i\ne j\}$ constitutes the set of negative pairs. With that, Equation \ref{eq:loss_instance} shows the cross-modal normalized temperature-scaled cross-entropy objective~\cite{chen2020simple} associated with the $i^{th}$ instance. Since $(v, a) \in \{(v_p, a_s), (v_s, a_p)\}$, in practice we optimize Equation \ref{eq:loss_cross} which aggregates over all available instances. 

\begin{equation}\label{eq:loss_instance}
    \ell_{i}(v,a) = -\log\Big(\frac{e^{((z^{i}_{v})^\intercal (z^{i}_{a}))/\tau}}{e^{((z^{i}_{v})^\intercal (z^{i}_{a}))/\tau} + \displaystyle\sum_{(z'_{v},z'_{a})\in \mathcal{N}_i}e^{((z'_{v})^\intercal (z'_{a}))/\tau}}\Big)
\end{equation}

\begin{equation}\label{eq:loss_cross}
    \mathcal{L} = \dfrac{1}{2} \sum_{i=1}^{B}\Big(\ell_{i}(v_p, a_s) + \ell_{i}(v_s, a_p)\Big)
\end{equation}

\begin{equation}\label{eq:loss_within}
    \mathcal{L}_v = \sum_{i=1}^{B}\ell_{i}(v_p, v_s), \quad \mathcal{L}_a = \sum_{i=1}^{B}\ell_{i}(a_p, a_s)
\end{equation}

Equation \ref{eq:loss_within} shows the within-modal variants of the loss function for video and audio modalities. Unless explicitly mentioned otherwise, we train our models from scratch and \textit{cross-modally}, \textit{i.e.}~we compute the contrastive loss between modalities as shown in Eq.~\ref{eq:loss_cross}. We do this based on the observation in our early experiments that, when training from scratch without tuning additional scaling parameters, the within-modal contrastive task is too easy comparatively and results in early convergence on the corresponding terms. This approach is also supported by prior literature~\cite{morgado2021audio}. Despite not directly optimizing for within-modal terms, we track $\mathcal{L}_v$ and $\mathcal{L}_a$ during self-supervised pretraining and observe that they diminish as a byproduct of minimizing $\mathcal{L}$. There are variants in our modeling where $\mathcal{L}_v$ and $\mathcal{L}_a$ are included in total loss function  (\textit{e.g} $\mathcal{L}+\lambda_v \mathcal{L}_v + \lambda_a \mathcal{L}_a$) which we'll discuss later in Sec.~\ref{sec:models}.

\subsection{Architecture}
As we seek to validate the effect of our data and training approach, we rely on standard backbone architectures. Our video model is a multi-scale vision transformer~\cite{fan2021multiscale}, specifically MViT-S, and our audio model follows a similar architecture except a slight modification to allow processing audio spectrograms as input. Note that we train all our models from scratch on our pretraining dataset detailed in Sec.~\ref{sec:pretraining_dataset}. We use a single (weight sharing) audio backbone which processes all audio spectrograms, regardless of language. As is common in contrastive learning, we use multi-layer perceptron (MLP) projection heads, one for each modality, to further reduce the dimensionality of representations during training, prior to computing the contrastive loss. These additional layers are discarded after pretraining.
\section{Experiments}
\label{sec:experiments}

\subsection{Downstream Tasks}
\paragraph{Audio Tasks and Benchmarks}
We evaluate on a diverse set of auditory tasks to probe the quality of our learned representations, taken from the HEAR~\cite{turian2022hear} challenge benchmark. We subselect tasks relevant to our hypotheses, and focus on those which use pooled (rather than temporally dense) representations. 

\textit{Sound and Scene Classification}: These tasks are firmly non-linguistic, and we hypothesize performance on them should benefit from de-emphasizing language in training. We include ESC-50~\cite{piczak2015environmental}, FSD50K~\cite{fonseca2021fsd50k}, and Vocal Imitations (VI)~\cite{kim2018vocal}. VI is a query-by-vocalization (QBV) task, however since it is based on AudioSet~\cite{gemmeke2017audio} ontology sound events, we place it in this category. \textit{Non-Semantic Speech}: Many non-semantic or \textit{paralinguistic} attributes of speech or vocal signals may be shared between languages, and such signals are important for a range of tasks. We include here CREMA-D~\cite{cao2014crema} for emotion recognition, GTZAN~\cite{tzanetakis1999} for music/speech discrimination, and LibriCount~\cite{stoter2018libricount} for speaker count estimation. We hypothesize performance should improve, if our scheme increases focus on non-linguistic speech attributes. \textit{Semantic Speech}: To probe a potential trade-off, we evaluate on semantic speech tasks. We consider keyword understanding as a proxy for speech recognition that uses pooled representations. To do so, we employ the \textit{full} version of Speech Commands~\cite{warden2018speech} implemented in HEAR~\cite{turian2022hear}. \textit{Language}: Another way to measure a possible trade-off is by evaluating how models perform on an audio-based language identification task, to see if features useful for this are preserved in learned representations. We include VoxLingua107 Top10~\cite{valk2021} for this reason.

\paragraph{Visual and Audiovisual Tasks}
\vspace{-1em}
We also evaluate the visual representations independently, and coordinated with, the auditory representations. Following recent work on representation learning from long-form content~\cite{chen2023movies2scenes}, we include the LVU~\cite{lvu2021} benchmark covering various aspects of long-form video understanding to our evaluation suite. LVU~\cite{lvu2021} contains small-scale tasks covering a wide range of aspects of long-form videos, including content understanding (\textit{relationship}, \textit{speaking style}, \textit{scene/place}), and movie metadata prediction (\textit{director}, \textit{genre}, \textit{writer}, \textit{movie release year}). Among the LVU tasks, we explore benefits and potential trade-offs using both visual and auditory representations. In general, we expect improvement except for \textit{speaking style}, where it is not \textit{a priori} clear whether de-emphasizing spoken words during pretraining is harmful for such a downstream task.

\paragraph{Evaluation}
\vspace{-1em}
Once the self-supervised pretraining is over, we discard the projection heads and use the backbone architectures to extract features from audio and video assets. Unless mentioned otherwise, we do spatio-temporal mean pooling on the output tensors in order to obtain a $d$-dimensional vector embedding for each data instance in the downstream tasks. We then train either an MLP or linear probe on these representations following the prescribed approaches in the relevant benchmarks. More implementation details can be found in the appendix.

\setlength{\tabcolsep}{4pt}

\begin{table}
    \small
    \centering
    \begin{tabular}{ccccccc}
    \toprule
     & \makecell{\#~data} & init. & ($\lambda_v$, $\lambda_a$) & \makecell{original\\language} & \makecell{avg.\\\#~dubs} & \makecell{dub \\ augment}\\
    \midrule
    
    \vIfmtZLV & 4.6M & rand. & (0,0) & \textbf{ESF} & 2.8 & \xmark\\
    \JnHtnpTW & 4.6M & rand. & (0,0) & \textbf{ESF} & 2.8 & \cmark\\
    \vIfmtZLVStageTwo & 4.6M & \vIfmtZLV & (0,0.2) & \textbf{ESF} & 2.8 & \xmark\\
    \JnHtnpTWStageTwo & 4.6M & \JnHtnpTW & (0,0.2) & \textbf{ESF} & 2.8 & \cmark\\ 
    \midrule
    
    \ErTWgScg & 11.8M & rand. & (0,0) & \textbf{EN} & 1.0 & \xmark\\
    \fHNHoNvT & 9.8M & rand. & (0,0) & $\mathbb{U}\backslash$\textbf{EN} & 0.2 & \xmark\\
    \sWdhUbZf & 19.4M & rand. & (0,0) & $\mathbb{U}$ & 0.6 & \xmark\\
    \uapnFuFg & 5.1M & \sWdhUbZf & (0,0) & $\mathbb{U}$ & 2.8 & \cmark\\
    \MZZCMvBJ & 5.1M & \sWdhUbZf & (0.2,0.2) & $\mathbb{U}$ & 2.8 & \cmark\\
    \midrule
    
    \PiShi & 19.4M & rand. & (0,0) & $\mathbb{U}$ & 0.6 & \cmark\\
    \KAraFs & 5.1M & \PiShi & (0.1,0.1) & $\mathbb{U}$ & 2.8 & \cmark\\
    \bottomrule
\end{tabular}
  \caption{Details of different pretraining model variants. Here, $\textbf{ESF}\coloneqq\{\textbf{EN}, \textbf{ES}, \textbf{FR}\}$ is denoting the union of three languages. $\mathbb{U}$ represents the universal set including all the seven languages.}
  \label{tab:models}
\end{table}

\begin{table*}
\setlength{\tabcolsep}{5.8pt}
  \footnotesize
  \centering
  \begin{tabular}{cccccccc|ccccccc}
    \toprule
     & \multicolumn{7}{c}{HEAR} & \multicolumn{7}{c}{LVU}\\
    \cmidrule(lr){2-8}\cmidrule(lr){9-15}
    & \makecell{ESC} & \makecell{LibCnt} & \makecell{CREMA} & \makecell{VI} & \makecell{FSD} & \makecell{Speech} & \makecell{VoxLng} & Director & Genre & Relation & Scene & Speak & Writer & Year \\    
    \midrule
    \vIfmtZLV & 77.20 & 67.29 & 59.52 & 10.37 & 44.52 & 74.83 & 27.16 & 44.86 & 54.42 & 36.59 & \textbf{45.12} & 42.86 & \textbf{38.10} & 41.84 \\
    \JnHtnpTW & 75.95 & 67.94 & 59.76 & 11.14 & 44.23 & 73.80 & 23.87 & 47.66 & 56.63 & 36.59 & 41.46 & 40.74 & 33.33 & 41.84 \\
    \vIfmtZLVStageTwo & 82.00 & 67.87 & \textbf{62.69} & 11.39 & 48.90 & \textbf{79.47} & \textbf{28.70} & \textbf{49.53} & 57.65 & 43.90 & 39.02 & 43.92 & 33.93 & 46.10 \\
    \JnHtnpTWStageTwo & \textbf{83.05} & \textbf{68.65} & 61.95 & \textbf{12.57} & \textbf{49.42} & 74.38 & 26.55 & 44.86 & \textbf{59.01} & \textbf{46.34} & \textbf{45.12} & \textbf{48.15} & 29.17 & \textbf{47.52} \\
    \midrule
    \ErTWgScg & 84.15 & 67.12 & 61.00 & \textbf{13.05} & 50.29 & 82.31 & 24.69 & 47.66 & 57.14 & 51.22 & 41.46 & 42.33 & 32.14 & \textbf{45.39} \\
    \fHNHoNvT & 82.00 & 67.10 & 61.98 & 11.86 & 49.07 & 82.90 & 28.09 & 42.99 & 55.95 & 48.78 & 42.68 & 47.62 & 30.36 & 44.68 \\
    \sWdhUbZf & \textbf{85.60} & 66.31 & 62.79 & 11.55 & \textbf{53.69} & \textbf{83.82} & \textbf{30.35} & 50.47 & \textbf{60.20} & 46.34 & 42.68 & 48.68 & 37.50 & \textbf{45.39} \\
    \uapnFuFg & 83.75 & 68.88 & 63.18 & 10.82 & 51.61 & 77.12 & 28.19 & \textbf{51.40} & 59.69 & \textbf{56.10} & 46.34 & \textbf{49.21} & \textbf{38.10} & 44.68 \\
    \MZZCMvBJ & 85.25 & \textbf{69.16} & \textbf{63.27} & 11.38 & 52.48 & 76.99 & 27.98 & \textbf{51.40} & 58.33 & 51.22 & \textbf{52.44} & 48.68 & 36.31 & \textbf{45.39} \\
    \midrule
    \PiShi & 84.10 & 67.57 & 63.70 & \textbf{12.12} & 51.96 & \textbf{81.88} & 29.42 & 42.99 & \textbf{58.84} & 48.78 & 46.34 & 41.27 & 38.69 & 41.13 \\
    \KAraFs & \textbf{85.50} & \textbf{68.90} & \textbf{64.28} & 11.90 & \textbf{52.55} & 77.14 & \textbf{29.94} & \textbf{48.60} & 57.65 & 48.78 & \textbf{51.22} & \textbf{50.79} & \textbf{39.88} & \textbf{49.65} \\
    \bottomrule
  \end{tabular}
  \caption{\textbf{Ablation results with audio.} All metrics are top-1 accuracy, except for FSD50K~\cite{fonseca2021fsd50k} and Vocal Imitation~\cite{kim2018vocal} (Mean Average Precision). We have followed the prescribed evaluation strategy from HEAR~\cite{turian2022hear} benchmark; training an MLP on frozen embeddings of the downstream tasks. For LVU~\cite{lvu2021}, we use the official data splits and train a linear probe. Results are shown on the test split where the best epoch to report is chosen based on the same metric on the validation set. All model variants obtained 100.0 top-1 accuracy on GTZAN, hence we did not include that task here. We denote the top performance(s) within each ablation group with \textbf{bold}. The HEAR~\cite{turian2022hear} tasks from left to right are ESC-50, LibriCount, CREMA-D, Vocal Imitation, FSD-50k, SpeechCommands (Full), and VoxLingua107 Top10.}
  \label{tab:ablation_audio}
\end{table*}

\begin{table*}
\setlength{\tabcolsep}{2.2pt}
  \footnotesize
  \centering
  \begin{tabular}{lcccccccc|lccccccc}
    \toprule
     & \multicolumn{7}{c}{HEAR} & & & \multicolumn{7}{c}{LVU}\\
    \cmidrule(lr){2-9}\cmidrule(lr){11-17}
    & \makecell{ESC} & \makecell{LibCnt} & \makecell{CREMA} & \makecell{VI} & \makecell{FSD} & \makecell{Speech} & \makecell{VoxLng} & \makecell{GTZAN} & & Director & Genre & Relation & Scene & Speak & Writer & Year \\    
    \midrule
    Bench~\cite{turian2022hear} & 96.65 & 78.53 & 75.21 & 22.69 & 65.48 & 97.79 & 72.02 & 99.23 & Obj Tr~\cite{lvu2021} & 58.90 & 56.10 & 54.70 & 60.00 & 40.30 & 35.10 & 40.60 \\
    Bench (SSL) & 80.50 & 78.53 & 75.21 & 18.48 & 50.88 & 96.87 & 71.40 & 96.86 & M2S~\cite{chen2023movies2scenes} & 70.90 & 55.90 & 71.20 & 68.20 & 42.20 & 53.70 & 57.80 \\
    GURA~\cite{wu2022efficacy}    & 74.35 & 68.34 & 75.21 & 18.48 & 41.32 & 94.68 & 71.40 & 93.59 & ViS4mer~\cite{islam2022long} & 62.61 & 54.71 & 57.14 & 67.44 & 40.79 & 48.80 & 44.75 \\
    PaSST~\cite{koutini2021efficient} & 94.75 & 66.01 & 61.04 & 18.20 & 64.09 & 63.87 & 25.93 & 97.69 & SCALE~\cite{sameni2023spatio} & 49.09 & 58.97 & 76.47 & 74.02 & 42.27 & 62.76 & 39.23 \\
    CLAP~\cite{elizalde2023clap} & 96.70 & 77.83 & 64.36 &   --  & 58.59 & 96.83 &  --  & 100.0 & STCA~\cite{diba2023spatio} & 66.70 & 56.62 & 59.25 & 69.15 & 41.62 & 52.93 & 53.30 \\
    \toprule
     \multicolumn{17}{c}{~~~~~~~~~~~~Ours}\\
    \midrule     
    \sWdhUbZf (A) & 85.60 & 66.31 & 62.79 & 11.55 & 53.69 & 83.82 & 30.35 & 100.0 & \sWdhUbZf (V) & 69.16 & 60.88 & 60.98 & 63.41 & 46.03 & 48.81 & 52.48 \\
    \uapnFuFg (A) & 83.75 & 68.88 & 63.18 & 10.82 & 51.61 & 77.12 & 28.19 & 100.0 & \uapnFuFg (V) & 67.29 & 61.73 & 60.98 & 65.85 & 47.62 & 41.67 & 55.32 \\
    \MZZCMvBJ (A) & 85.25 & 69.16 & 63.27 & 11.38 & 52.48 & 76.99 & 27.98 & 100.0 & \MZZCMvBJ (V) & 69.16 & 64.29 & 58.54 & 64.63 & 46.03 & 41.07 & 52.48 \\  
    \bottomrule
  \end{tabular}
  \caption{State-of-the-art results across HEAR~\cite{turian2022hear} (adding GTZAN Music/Speech) and LVU~\cite{lvu2021} tasks we evaluate on. On HEAR, we compare to (1) the best result on each task, on the HEAR leaderboard, (2) same as (1) but considering only self-supervised models, (3) GURA Fuse HuBERT~\cite{wu2022efficacy}, the best performer on average, (4) CP-JKU PaSST 2lvl+mel~\cite{koutini2021efficient}, the strongest average performer after the GURA models, (5) the recent CLAP model~\cite{elizalde2023clap}. On LVU, we compare to the Object Transformer from the original LVU paper~\cite{lvu2021}, along with recent advances: ViS4mer~\cite{islam2022long}, the SVT SCALE model~\cite{sameni2023spatio}, STCA~\cite{diba2023spatio}, and Movies2Scenes~\cite{chen2023movies2scenes}. Movies2Scenes uses movie metadata, which introduces task-specific supervision. When reporting our results, (A) indicates audio representations only, and (V) means video representations only.}
  \label{tab:sota}
\end{table*}

\begin{table}
  \footnotesize
  \centering
  \begin{tabular}{cccccccc}
    \toprule    
     & Director & Genre & Relation & Scene & Speak & Writer & Year \\
    \midrule            
    \vIfmtZLV & 53.27 & 54.59 & 43.90 & 52.44 & 34.39 & 36.90 & 42.55 \\
    \JnHtnpTW & 53.27 & 55.44 & 41.46 & 50.00 & \textbf{41.27} & 35.12 & 42.55 \\
    \vIfmtZLVStageTwo & 57.01 & \textbf{57.48} & \textbf{46.34} & \textbf{57.32} & 39.68 & \textbf{38.69} & 46.10 \\
    \JnHtnpTWStageTwo & \textbf{63.55} & \textbf{57.48} & 36.59 & 53.66 & 36.51 & 33.93 & \textbf{47.52} \\
    \midrule
    \ErTWgScg & 60.75 & 55.78 & \textbf{48.78} & 53.66 & 38.10 & 35.71 & 42.55 \\
    \fHNHoNvT & 54.21 & 57.65 & 46.34 & 51.22 & 37.04 & \textbf{38.69} & 44.68 \\
    \sWdhUbZf & \textbf{65.42} & 57.48 & 41.46 & 53.66 & 39.68 & 38.10 & 45.39 \\
    \uapnFuFg & 62.62 & \textbf{58.50} & 36.59 & \textbf{59.76} & \textbf{43.39} & 35.12 & 46.81 \\
    \MZZCMvBJ & 62.62 & 58.16 & 43.90 & \textbf{59.76} & 39.15 & 37.50 & \textbf{49.65} \\
    \midrule
    \PiShi & \textbf{63.55} & 55.10 & 43.90 & 57.32 & \textbf{40.74} & \textbf{39.29} & \textbf{45.39} \\
    \KAraFs & 61.68 & \textbf{56.63} & \textbf{46.34} & \textbf{60.98} & 40.21 & 36.90 & 43.97 \\
    \bottomrule    
  \end{tabular}
  \caption{\textbf{Ablation results with video}. All metrics are top-1 accuracy. We have followed prescribed data split from LVU benchmark and trained a linear probe on frozen \textbf{video} embeddings of the downstream tasks. Results are shown on the test split where the best epoch to report is chosen based on the validation set. We denote the top performance within each ablation group with \textbf{bold}.}
  \label{tab:lvu_video}
\end{table}

\subsection{Models}
\label{sec:models}

In total, we train \textbf{11} model variants, detailed in Table~\ref{tab:models}, and evaluate them on \textbf{15} different tasks across audio and video modalities.

\textbf{\textcolor{Acolor}{ First (A)}} group of model variants demonstrates a small-scale multilingual pretraining regime, as a first study of the impact of dub-augmentation. We sample English (\textbf{EN}), Spanish (\textbf{ES}), or French (\textbf{FR}) titles which have at least one dub available, so we can systematically study the effect of dub-augmentation. For each title, we sample dubs from \textit{all} seven total languages. \vIfmtZLVStageTwo and \JnHtnpTWStageTwo variants incorporate an explicit within-modal term, \textit{i.e} $\mathcal{L}_a$. We hypothesize that, with dub-augmentation, $\lambda_a > 0$ may yield a broader gap on linguistic and language identification tasks. This is because the optimization explicitly maximizes the similarity of audio embeddings that are only different in their spoken language, rather than just implicitly through $\mathcal{L}$. Importantly, the total number of pretraining steps is the same for \vIfmtZLVStageTwo and \JnHtnpTWStageTwo, similarly when one compares \vIfmtZLV and \JnHtnpTW.

\textbf{\textcolor{Bcolor}{ Second (B)}} group of model variants aims at understanding the impact of data scale and language diversity. We approximately double the number of pretraining instances compared to experiments in group \textbf{\textcolor{Acolor}{A}} and study whether this leads to higher quality representations. This is important since self-supervised pretraining is computationally expensive and it is not clear \textit{a priori} if bigger and more diverse pretraining data necessarily leads to better models. \sWdhUbZf is trained on all pretraining instances including all languages to test the limit of multilingual pretraining \textit{without} dub-augmentation. By comparing \uapnFuFg and \MZZCMvBJ, we hope to shed light on the behavior of the within-modal objective function which the latter uses.

\textbf{\textcolor{Ccolor}{ Third (C)}} group of experiments explore the impact of deeper architectures, namely MViT-B~\cite{fan2021multiscale} (vs MViT-S~\cite{fan2021multiscale} as our default). We keep the data scale and diversity the same as in the \sWdhUbZf, \uapnFuFg and \MZZCMvBJ variants. Similarly to these, here we initially train on the entire data, then fine-tune from the final checkpoint of \PiShi only on a subset of titles which have more than one audio tracks. This ensures that dub-augmentation is present in every optimization step of \KAraFs.

We are now set to comprehensively study various aspects of multilingual and multimodal representation learning, thanks to a wide variety of pretrained models and downstream tasks across audio and video modalities.

\subsection{Ablation Study}

\paragraph{Does dub-augmented pretraining help?}
To address this, we start by looking at the \textbf{\textcolor{Acolor}{first (A)}} group of model variants in Table~\ref{tab:ablation_audio}. We've hypothesized that dub-augmentation should improve the performance on sound/scene classification and non-semantic speech tasks. On the HEAR~\cite{turian2022hear} benchmark, with the exception of CREMA-D~\cite{cao2014crema}, our quantitative results confirm this. LVU~\cite{lvu2021} tasks are also considered non-linguistic and Table~\ref{tab:ablation_audio} shows that, in most of them, dub-augmented variants lead to large performance gains over their baseline counterparts. Our second hypothesis was that dub-augmentation should impact linguistic and language identification tasks as it aims at diminishing the influence of spoken words in audio representations. Indeed, we can see \JnHtnpTWStageTwo which utilizes dub-augmentation is underperforming \vIfmtZLVStageTwo on Speech Commands and VoxLingua. Table~\ref{tab:ablation_audio} also suggests that dub-augmentation benefits from within-modal objective \textit{i.e.} $\mathcal{L}_a$, and for this approach to be effective, we actually need as expected, sufficient number of instances with alternative audio tracks during pretraining.

\paragraph{Can dub-augmented models still recognize language and conduct linguistic tasks?}
\vspace{-1em}
Results shown in Table~\ref{tab:ablation_audio} on VoxLingua demonstrate that enforcing dub-augmentation in both small (\textbf{\textcolor{Acolor}{A}} variants) and large-scale (\textbf{\textcolor{Bcolor}{B}} variants) regimes clearly affects language identification performance. We measure this by comparing \JnHtnpTW vs. \vIfmtZLV, or \uapnFuFg vs. \sWdhUbZf. We observe similar behavior for Speech Commands~\cite{warden2018speech}, our proxy for linguistic performance implemented as keyword spotting. However, in both cases, the degradation is not large enough to prevent dub-augmented models from recognizing language or conducting linguistic tasks. We hypothesized this modeling trade-off, \textit{i.e.} that while performance might reduce, the significance of this would be limited.

\paragraph{Is the quality of video representations impacted?}
\vspace{-1em}
To answer this, we look at Table~\ref{tab:lvu_video} where LVU~\cite{lvu2021} tasks are evaluated via a linear probe on frozen video embeddings. In the small-scale pretraining regime, we observe a mixed pattern where dub-augmented variants, \textit{i.e.} \JnHtnpTW and \JnHtnpTWStageTwo, outperform their counterparts in 3 tasks ("Director", "Speaking Way", and "Year") while being either worse or on par on the rest. In the large-scale pretraining regime, we see a more clear trend where \uapnFuFg and \MZZCMvBJ show improvements over \sWdhUbZf in 5 out of 7 LVU tasks demonstrating that on a diverse evaluation set, dub-augmented pretraining is overall helpful to even video-only tasks.

\paragraph{How does language diversity influence pretraining?}
\vspace{-1em}
Properly addressing this research question demands a closer look at \ErTWgScg, \fHNHoNvT, and \sWdhUbZf. It is worth reiterating that despite a different number of pretraining instances (see Table~\ref{tab:models}), we have trained all three of these model variants with approximately the same number of gradient optimization steps to establish a fair comparison.
In general, across both audio (ref. Table~\ref{tab:ablation_audio}) and video (ref. Table~\ref{tab:lvu_video}) we observe performance gains when we maximize language variation (ref. \sWdhUbZf). However, the inclusion of English (\textbf{EN}) language titles, as our most dominant original language (see Fig~\ref{fig:dataset_stat}), during pretraining seems to be crucial. Table~\ref{tab:ablation_audio} illustrates a clear pattern for VoxLingua~\cite{valk2021} and Speech Commands~\cite{warden2018speech}, where greater language diversity during pretraining leads to significant gains \textit{e.g.} absolute 5.6\% on VoxLingua~\cite{valk2021}.

\paragraph{Is a deeper architecture better?}
\vspace{-1em}
For each task in Tables~\ref{tab:ablation_audio} and \ref{tab:lvu_video}, we can compare the strongest \textbf{\textcolor{Bcolor}{B}} model variants against \textbf{\textcolor{Ccolor}{C}} variants. With a few exceptions, our quantitative results do not indicate that using MViT-B~\cite{fan2021multiscale} with $\sim$40\% more parameters provides a meaningful boost over the smaller MViT-S~\cite{fan2021multiscale} to justify the significant additional computation during pretraining. We acknowledge that this conclusion might not have held if downstream tasks where evaluated by fine-tuning (instead of linear/MLP probing), especially for large-scale tasks in HEAR~\cite{turian2022hear}.

\paragraph{Additional Experiments}
\vspace{-1em}
In the appendix, we provide additional results on a small dubbed audiovisual dataset with matched smaller backbone architectures, where we have exact parity between four languages (over 700 \textbf{EN} titles with all of \textbf{ES}, \textbf{FR}, and \textbf{JA} available). We also compare to a speech-removal strategy, where we source-separate the full dataset and remove the speech part as an alternate strategy for de-emphasizing the speech. Since we have language parity, we also evaluate "bilingual" models with specific dub-augmentation pairs (e.g. \textbf{EN}+\textbf{ES}). These results show systematically that dub-augmented training is beneficial even in this smaller-scale setup, that it outperforms the speech removal strategy, and that multilingual models (with multiple dubs, randomly sampled as in our main results here) can add further robustness.

\subsection{Comparison with State-of-the-Art}

\paragraph{HEAR}
Table~\ref{tab:sota} compares our results to several strong results on HEAR~\cite{turian2022hear} tasks. On ESC-50, FSD50K, and GTZAN Music/Speech, our results beat the top self-supervised result on the HEAR Leaderboard and at least one more result. On most tasks (except Vocal Imitation), we beat at least one of the models, showing robustness across these different tasks.

\paragraph{LVU}
\vspace{-1em}
Also in Table~\ref{tab:sota}, we compare our strongest models with state-of-the-art results on 7 LVU~\cite{lvu2021} tasks. Our models achieve new state-of-the-art performance on the \textit{Genre} and \textit{Speak} tasks, showing substantial improvements over prior results. Without considering Movies2Scenes~\cite{chen2023movies2scenes}, which uses movie metadata, we also get state-of-the-art results on \textit{Director} and \textit{Year} (4/7 total). On the remaining tasks, our results are highly competitive. This demonstrates that models pretrained on our dataset with dub-augmentation can match or exceed the performance of the best available models on a diverse range of video understanding benchmarks. Overall, these results highlight the effectiveness of our approach.
\section{Synthetic Counterfactual Pairs}
\label{sec:synthetic_dataset}

\begin{figure*}[!htb]
    \centering
    \includegraphics[width=\textwidth]{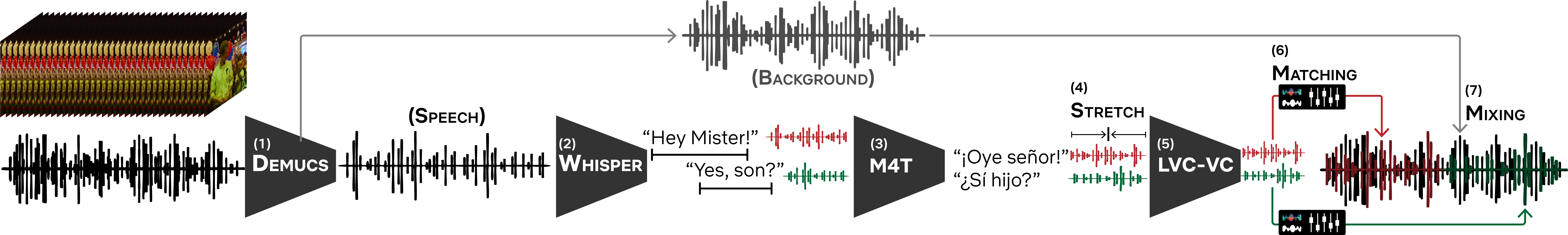}
    \caption{Pipeline to produce the synthetic counterfactual pairs.}
    \label{fig:synthetic_dataset}
\end{figure*}

To encourage the study of counterfactual pairs in audiovisual representation learning, we propose a modular pipeline, shown in Fig.~\ref{fig:synthetic_dataset}, for simulating dub-like counterfactual pairs that are similar to the one-to-many audio-visual distribution from our pretraining data on arbitrary target clips. The proposed pipeline, while being limited in terms of the synthetic quality, serves as a simple tool to alleviate the data constraint for the research community when conducting a similar study.  

The steps are (1) Isolate speech from background sounds using Demucs~\cite{defossez2019demucs}, (2) Transcribe and segment the speech using Whisper~\cite{radford2023robust}, producing timestamped segments (3) Translate speech (or, optionally, text) into the target language(s) with SeamlessM4T~\cite{barrault2023seamlessm4t} (4) Align translations to original segments using stretching (5) Convert voices to match original actors' using LVC-VC~\cite{kang2022end} (6) Loudness-normalize and EQ-match the output with the original using Pyloudnorm~\cite{steinmetz2021pyloudnorm} and matchering\footnote{\href{https://github.com/sergree/matchering}{https://github.com/sergree/matchering}} (7) re-place segments into their original locations, remix with background audio, and mux with original videos. The pipeline also implements other intermediate steps, such as resampling, to bridge between the main steps.

As a proof-of-concept resource for the community, we use this pipeline to produce a multilingual version of LVU~\cite{lvu2021}. LVU-M demonstrates the feasibility of generating counterfactual data at scale. We will open-source the pipeline to enable creating such \textit{``looking similar, sounding different"} datasets. We also hope that future advancements can improve the quality and enable deeper research of such data structure.
\section{Conclusion}
\label{sec:conclusion}
In this work, we introduced the \textit{looking similar, while sounding different} problem, wherein perceptually similar scenes can have different speech content. We showed we can leverage a similarly structured counterfactual data source, dubbed movies, to improve audiovisual representation learning in a well-established cross-modal contrastive learning scheme. Our experiments with a large pretraining dataset of movies and television shows demonstrated that this improves performance across a range of auditory and audiovisual tasks. Dub-augmented training is, as such, a scalable and effective approach for learning more robust audiovisual representations without supervision.

{
    \small
    \bibliographystyle{ieeenat_fullname}
    \bibliography{main}
}

\addtocontents{toc}{\protect\setcounter{tocdepth}{-1}}

\clearpage
\setcounter{page}{1}
\appendix

\renewcommand{\thefigure}{\thesection.\arabic{figure}}
\renewcommand{\thetable}{\thesection.\arabic{table}}

\setcounter{figure}{0}
\setcounter{table}{0}

\maketitlesupplementary

\begin{abstract}
    In this appendix, we present information about our pretraining procedures and results from additional experiments. We also showcase examples of our synthetic data. This document is laid out as shown below.
    {\hypersetup{linkcolor=blue}\color{blue}\small\tableofcontents}
\end{abstract}

\addtocontents{toc}{\protect\setcounter{tocdepth}{3}}

\section{Pretraining Details}
\begin{figure}[!htb]
    \centering
    \includegraphics[width=\columnwidth]{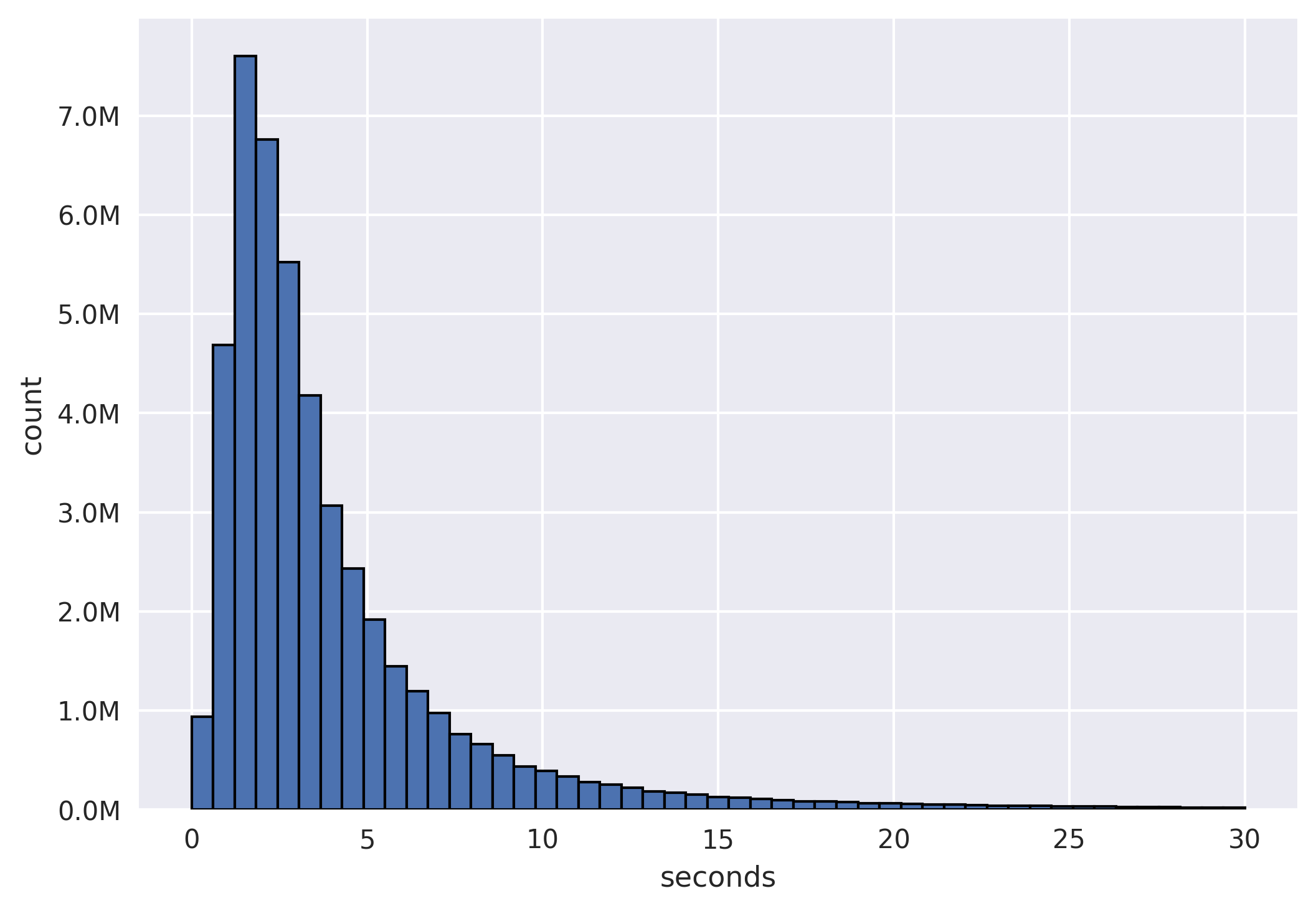}
    \caption{Distribution of shot lengths observed in our dataset.}
    \label{fig:clip_length}
\end{figure}

\paragraph{Data Preprocessing}
We temporally segment long-form content into shots (camera changes). Fig.~\ref{fig:clip_length} shows the distribution of shot lengths. We ignore shots that are shorter than 3 and longer than 12 seconds. The former constraint is to make sure the snippet is long enough for our models, while the latter is to improve training throughput. The total number of shots in each pretraining setting is shown in Table~\ref{tab:models} under the column \#data. When creating a minibatch during pretraining, we ensure that $\frac{1}{8}$ of each batch comes from the same long-form content source (e.g. the same movie) to create hard negatives. The process of generating quadruple training instances ($v_p$, $a_p$, $v_s$, $a_s$) is as follows:
\begin{enumerate}
    \item Given a title, randomly pick a shot.
    \item Temporal jitter: randomly select two 3-second temporal windows. These two snippets, derived from the same shot, are our primary and secondary instances. For the secondary instance, the language of the audio is different from the one in the primary instance, if an alternate audio track (i.e. dub) is available.
    \item For each pair of audio and video: \\
    \textbf{Video}: Resample to 25 fps, uniformly sample 16 frames, randomly scale the shorter side of video within the range of 256-320, then perform a random crop of 224x224. \\
    \textbf{Audio}: Resample to 48kHz, convert audio to mel-spectrogram (n\_fft=1024, hop\_length=501, num\_mels=96), convert to the decibel scale, and apply time and frequency masking with maximum value of 50 percent of the corresponding axis.
\end{enumerate}

\paragraph{Model and Pretraining Hyperparameters}
The MLP projection heads have an output dimensionality of 512. The latent embeddings ($z$) are L2 normalized prior to computing the loss. The temperature factor $\tau$ in the objective function is set to 0.07. We use the AdamW optimizer~\cite{loshchilov2017decoupled} with a learning rate of 3e-4, and weigh decay of 5e-2. We train for 12 epochs on 32 NVIDIA A100 GPUs, with a batch size of 64 per GPU, using a half-cosine learning rate annealing which kicks off after 2 warm-up epochs.

\section{Additional Experiments}
\subsection{Results on Action Recognition}
We report results on \textbf{UCF101}~\cite{soomro2012ucf101} and \textbf{HMDB51}~\cite{kuehne2011hmdb}, well-known benchmarks, to assess the video-only performance of our models, shown in Table~\ref{tab:video_performance}. Performance between our model variants is comparable, showing that the dub-augmented training does not necessarily decrease video-only performance. Additionally, we compare to recent state-of-the-art results which, like us, do not use fine-tuning. Note that these results use linear probes, vs. our MLP probes which were derived from a grid search over probing strategies. Nevertheless, the fact that we significantly beat these results without fine-tuning ($>$12\% absolute) demonstrates the value of our learned representations.

\begin{table}
\centering
\begin{tabular}{lcc}
\toprule
\textbf{Model} & \textbf{UCF101}~\cite{soomro2012ucf101} & \textbf{HMDB51}~\cite{kuehne2011hmdb} \\
\midrule
\sWdhUbZf & 88.90 & 69.35 \\
\uapnFuFg & 88.20 & 68.91 \\
\MZZCMvBJ & 87.99 & 69.43 \\
\midrule
FIMA~\cite{zhu2023fine} & 76.40 & 47.30 \\
FAME~\cite{ding2022motion} & 72.20 & 42.20 \\
\bottomrule
\end{tabular}
\caption{Performance of video models on UCF101~\cite{soomro2012ucf101} and HMDB51~\cite{kuehne2011hmdb} datasets, comparing with recent results that \textit{do not} involve fine-tuning.}
\label{tab:video_performance}
\end{table}

\subsection{VGGSound Results}
We report results on \textbf{VGGSound}~\cite{chen2020vggsound}, an audiovisual benchmark on which we focus on audio results, shown in Table~\ref{tab:audio_performance}. Once again, performance between our model variants is comparable, and our results are competitive with recent state-of-the-art results which don't use fine-tuning.

\begin{table}
\centering
\begin{tabular}{llc}
\toprule
\textbf{Model} & \textbf{VGGSound} \\
\midrule
\sWdhUbZf & 43.49 \\
\uapnFuFg & 41.95 \\
\MZZCMvBJ & 42.96 \\
\midrule
LAION-CLAP~\cite{wu2023large} & 46.20 \\
BLAT~\cite{xu2023blat} & 42.90 \\
\toprule
\end{tabular}
\caption{Performance of audio models on the VGGSound~\cite{chen2020vggsound} dataset, comparing with recent results that \textit{do not} involve fine-tuning on the downstream dataset. The LAION-CLAP result reported uses keyword-to-caption augmentation.}
\label{tab:audio_performance}
\end{table}

\subsection{Controlled Dataset and Models}
In this section, we discuss the methods and results from a smaller-scale, more controlled set of experiments. The pretraining dataset consists of 748 movies, about 1300 video-hours of content. Each movie contains a video track, as well as four audio tracks: English (\textbf{EN}) as the primary language, and three dubbed versions, Spanish (\textbf{ES}), French (\textbf{FR}), and Japanese (\textbf{JA}), all languages for which we find dubs are relatively commonly available. Having multiple dub options allows us to investigate trade-offs between secondary languages, and whether ``multilingual'' models might further strengthen performance.

The video model is a medium X3D~\cite{feichtenhofer2020x3d}, which is an efficient ResNet-based model. Our audio model is an Acoustic ResNet50~\cite{xiao2020audiovisual}, which takes audio spectrograms as input. Both models output 1024-dimensional representations per clip. We share backbone weights (i.e. Acoustic ResNet50) across audio variants with primary and secondary (dubbed) languages. We do not share MLP weights for primary vs. secondary audio, to allow for more flexibility. As in our primary experiments, we mainly train these models \textit{cross-modally}, i.e.~we compute the contrastive cost between modalities.

We train these models on 4 A100 GPUs for 10 epochs with a batch size of 26 per GPU. We use a negative sampling parameter $k$ (samples drawn from the same movie as the positive clip), which we set to 12 per GPU. We use the AdamW optimizer~\cite{loshchilov2017decoupled} with $\beta$=(0.9, 0.999), a learning rate of 0.001, weight decay of 0.05, and a cosine learning rate schedule with a half-epoch warmup.

In all, we compare the following model variants in these smaller-scale, more controlled, experiments:

\begin{enumerate}
    \item \textbf{\textit{Monolingual}} (\textbf{EN}): In this baseline, we consider models trained with two differently-augmented primary (English) audio treated as ``primary'' and ``secondary'' ($a_p$=\textbf{EN}; $a_s$=\textbf{EN}) audio respectively. This is to account for any possible effect of two augmentations per seen sample, as occurs for the dub-augmented cases, although it does not modify the data distribution. This is a SimCLR-based setup, with two audio paths each contrasted with video.
    \item \textbf{\textit{Bilingual}} (\textbf{ES}, \textbf{FR}, \textbf{JA}): We introduce one secondary audio at a time to explore the dub-augmented training hypothesis ($a_p$=\textbf{EN}; $a_s$= \textbf{ES}~\textsc{or} \textbf{FR}~\textsc{or} \textbf{JA}).
    \item \textbf{\textit{Multilingual}} (+\textbf{EFJ}): Here, we effectively randomly select a secondary audio from the given list (Spanish, French, and Japanese) per batch ($a_p$=\textbf{EN}; $a_s \in_R$ \{\textbf{ES}, \textbf{FR}, \textbf{JA}\}). The order of samples is randomized, so in practice we simply circle through the list round-robin. We aim to explore whether there are additional benefits or drawbacks to having more than one secondary audio.
    \item \textbf{\textit{No-Speech}} (\textcolor{red}{\textbf{SEP}}): We establish another baseline where the speech is separated and we only train on video + non-speech audio. This allows us to examine whether simply removing the speech is enough for a performance gain on non-speech-focused tasks. We use the pretrained Hybrid Demucs v3 model~\cite{defossez2019demucs} to separate the vocal from the rest, mixing the other stems back together. There is no secondary audio here ($a_p$=\textbf{EN} $_\textrm{\textcolor{red}{SEP}}$). Note that this variant is trained with 44.1kHz audio, as this is the input and output sample rate for the Demucs models. Although Demucs is trained for music separation, we find that it works well on speech in practice on our dataset. We use the default (\texttt{mdx\_extra\_q}) pretrained model.
    \item \textbf{\textit{Audio-Only}} (Monolingual: \textcolor{brown}{\textbf{AUD}}, and Multilingual: \textcolor{brown}{\textbf{AUD}}{\tiny+\textbf{EFJ}}): Finally, we examine two audio-only models. The data is similar to the \textbf{\textit{monolingual}} and \textbf{\textit{multilingual}} setups, except without video. The objective function is now \textit{within-modal}, between the two audio clips. The monolingual version represents standard audio contrastive training with two augmented copies. These models cannot work on visual or audiovisual tasks, but here we seek to evaluate whether and how much dub-augmented training contributes improvements in the absence of video.
\end{enumerate}

\subsubsection{Evaluation}
\paragraph{Evaluation Tasks}
Beyond the HEAR~\cite{turian2022hear} tasks used in our main experiments, we include results from additional audio tasks to this controlled setup to gain a more complete picture in the controlled setup. First, we add audio tasks from HARES~\cite{wang2022towards}; specifically, TUT18~\cite{mesaros2018multi} for acoustic scene recognition, Fluent Speech Commands~\cite{lugosch2019speech} for speech command recognition, and VoxForge~\cite{maclean2018voxforge} for language identification, complementing existing HEAR tasks. As in the appendix for our main results, we include the video-only action recognition tasks HMDB51~\cite{kuehne2011hmdb} and UCF101~\cite{soomro2012ucf101}. Finally, we add an \textit{audiovisual} task (VGGSound~\cite{chen2020vggsound}) to facilitate a better comparison with~\textcolor{red}{\textbf{SEP}}, since this baseline sees no speech altogether. We hypothesize that \textcolor{red}{\textbf{SEP}} will be a strong performer in some cases, but that dub-augmented models will be stronger in general as they preserve the audiovisual relationship between speech actions visually occurring and sounding.

For the visual and audiovisual tasks, we train the linear probes for 200 epochs using Stochastic Gradient Descent and a learning rate of 0.2 following a cosine schedule. We train on 2 A10 GPUs with a total batch size of 1024. For HEAR tasks, we use the provided API's strategy and the 48kHz data. For HARES tasks, we follow the authors' specifications~\cite{wang2022towards}: in general, with 400K training steps and a learning rate schedule consisting of 5K linear warmup steps and a cosine decay for the rest (max. learning rate of 0.0002, with the Adam~\cite{kingma2014adam} optimizer). We train on 2 GPUs with a total batch size of 64. In all relevant cases, we duplicate mono audio to the second channel to form a pseudo-stereo input to match our model's architecture.

\subsubsection{Results}

\begin{table*}[!htb]
\centering
\begingroup\small
\begin{tabular}{lll|lll|lllll}
  \toprule \multicolumn{3}{c}{} & \multicolumn{3}{c}{\textbf{Baselines} (SimCLR)} & \multicolumn{5}{c}{\textbf{Dub-Augmented}} \\
 \cmidrule(l){4-6} \cmidrule(l){7-11}
  & Task & M & \textcolor{black}{AV} & \textcolor{red}{SEP} & \textcolor{brown}{A} & \textbf{ES} & \textbf{FR} & \textbf{JA} & \textbf{EFJ} & \textcolor{brown}{A}\tiny{+\textbf{EFJ}} \\ 
  \midrule
 \midrule
\multirow{5}{*}{\rotatebox[origin=c]{90}{\textbf{Snd/Scn}}} & ESC-50
\cite{piczak2015esc} & A & .527\scriptsize{±.012} & .570\scriptsize{±.028} & \textcolor{gray}{.220\scriptsize{±.027}} & .580\scriptsize{±.019} & .575\scriptsize{±.031} & \textbf{.590\scriptsize{±.036}} & .587\scriptsize{±.009} & .550\scriptsize{±.026} \\ 
   & FSD50K
\cite{fonseca2021fsd50k} & A & .296 & .307 & \textcolor{gray}{.109} & \textbf{.317} & .313 & .311 & .313 & .277 \\ 
   & TUT18
\cite{mesaros2018multi} & A & .853 & .857 & \textcolor{gray}{.682} & \textbf{.884} & .881 & .849 & .867 & .801 \\ 
   & VocalImitation
\cite{kim2018vocal} & A & .042 & .051 & \textcolor{gray}{.022} & .045 & .047 & .045 & .050\scriptsize{±.006} & \textbf{.055} \\ 
   & VGGSound
\cite{chen2020vggsound} & AV & .303 & \textcolor{gray}{.287} & — & \textbf{.323} & .314 & .314 & .311 & — \\ 
   \midrule
\multirow{3}{*}{\rotatebox[origin=c]{90}{\textbf{NonSem}}} & CREMA-D
\cite{cao2014crema} & A & .514\scriptsize{±.012} & .489\scriptsize{±.009} & \textcolor{gray}{.354\scriptsize{±.022}} & .528\scriptsize{±.009} & .540 & .520\scriptsize{±.010} & \textbf{.548\scriptsize{±.012}} & .530\scriptsize{±.011} \\ 
   & GTZAN Mus/Sp
\cite{tzanetakis1999} & A & .954\scriptsize{±.054} & .931\scriptsize{±.099} & \textcolor{gray}{.866\scriptsize{±.119}} & .946\scriptsize{±.082} & .891\scriptsize{±.142} & .931\scriptsize{±.092} & \textbf{.969\scriptsize{±.054}} & .954\scriptsize{±.054} \\ 
   & LibriCount
\cite{stoter2018libricount} & A & .654\scriptsize{±.026} & .608\scriptsize{±.016} & \textcolor{gray}{.505\scriptsize{±.014}} & .671\scriptsize{±.025} & \textbf{.706\scriptsize{±.021}} & .681\scriptsize{±.016} & .676\scriptsize{±.013} & .678\scriptsize{±.022} \\ 
   \bottomrule
\end{tabular}
\endgroup
\vspace{0.2em}
\caption{\textbf{Controlled experiments evaluation results.} All metrics are top-1 accuracy, except FSD50K~\cite{fonseca2021fsd50k} and VocalImitation~\cite{kim2018vocal} (Mean Average Precision). Results in \textbf{bold} indicate the highest score, and in \textcolor{gray}{gray} indicate the lowest. The task types are \textbf{Snd/Scn} = Sound/Scene Classification and \textbf{NonSem} = Non-Semantic Speech.}
\label{tab:results}
\vspace{-1.4em}
\end{table*}

In total, we trained 8 different model variants and evaluated them on 15 different tasks. Table~\ref{tab:results} shows our main tasks on which we hypothesized improvement (N=8), grouped by modality and task type.

\paragraph{Does dub-augmented pretraining help?}
For all tasks in Table~\ref{tab:results}, one or more dub-augmented models outperform the monolingual \textbf{EN}~model. In 6/8 tasks, \textit{all} dub-augmented variants outperform \textbf{EN}, except for the two easiest tasks (TUT18 and GTZAN). We hypothesized this outcome for the sound and scene classification tasks, where we consistently observe substantial gains, as well as the non-semantic speech tasks. This supports the results in the main paper.

\paragraph{Is the improvement due only to de-emphasizing speech?}
We examine the source-separated version to address this question, since it offers the extreme case where the speech is removed altogether (as much as possible). The source-separated variant presents a strong baseline on the sound/scene classification tasks, despite mostly being outperformed by one or more dub-augmented models. We expect this is due to re-focusing on non-speech elements. However, despite strong performance in these cases, this variant has drawbacks. First, it results in lower performance than all other models on VGGSound (audiovisual classification) and both visual tasks (shown in the trade-off results in Table~\ref{tab:tradeoffs_results}). We suspect this is because there is a clear discrepancy between the auditory and visual channels in the source-separated version, i.e. speech. When a person is speaking, and there is little or no speech content in the auditory stream accompanying the visual, this may act as a confounder for coordinating the two representations. Note that \textit{People} is a large category in VGGSound\footnote{\href{https://www.robots.ox.ac.uk/~vgg/data/vggsound/}{www.robots.ox.ac.uk/~vgg/data/vggsound}}.

Second, \textcolor{red}{\textbf{SEP}}~significantly underperforms on non-semantic speech tasks and (in Table~\ref{tab:tradeoffs_results}) language identification, with the exception of GTZAN which we find is an easier task in general. This intuitively makes sense: this variant does not see speech, effectively, and performs lower than the monolingual variant as well. These results illustrate a trade-off: source-separation as a preprocessing method, in addition to being very computationally expensive and weakening the self-supervision assumption (by dependence on a third-party supervised model), results in poor performance on paralinguistic tasks, which require attention to aspects of speech beyond language.

\paragraph{Are more languages better?}
Given the strength of dub-augmented training, we ask whether introducing more languages into the mix improves performance further. Our results don't indicate this to be the case, but note that in Table~\ref{tab:results}, the \textbf{EFJ}~model is least commonly the lowest-performing dub-augmented variant (1/8 tasks). Additionally, the multilingual variant performs well on 2/3 non-semantic speech tasks. Even though paralinguistic features can vary by language, commonalities exist that may be useful and many practical scenarios could benefit from diverse examples. The robustness of the multilingual model suggests that it could be a reasonable default choice assuming little knowledge about the downstream tasks, and we use a similar multilingual approach in our larger scale experiments in the main paper.

\paragraph{Is dub-augmentation beneficial even without video?}
The \textcolor{brown}{A}$_\textrm{+\textbf{EFJ}}$~variant always outperforms the \textcolor{brown}{\textbf{AUD}}~model (including on all audio tasks we examine later for trade-offs, shown in Table~\ref{tab:tradeoffs_results}). \textcolor{brown}{\textbf{AUD}}~is the weakest performer on all relevant tasks, indicating the benefits of cross-modal training. Additionally, on some tasks, the multilingual variant comes close to or even outperforms (as in on VocalImitation) the cross-modal variants. Of course, this variant cannot work on visual or multimodal tasks, and still largely underperforms the multimodal dub-augmented models, but it demonstrates the significant value of even unimodal dub-augmented training.

\subsubsection{Exploring Trade-Offs}
Results on the 7 tasks in Table~\ref{tab:tradeoffs_results} help us evaluate possible trade-offs in the smaller-scale and controlled setup, to complement the previous results.

\paragraph{Can dub-augmented models still recognize language?}
The dub-augmented variants generally perform similarly or slightly worse on VoxLingua~\cite{valk2021} but appear to do better on VoxForge~\cite{maclean2018voxforge}, both language identification tasks. The latter is a large-scale user-submitted dataset, which may have different auditory characteristics from the former as a result. Taking these results together, we expect that the dub-augmented models are able to retain information useful for language identification in their pre-MLP features. It is possible that more general auditory features, which do not encode speech semantics, are still discriminative in these tasks.

\paragraph{Are they discriminative between spoken words?}
As in our results from the main paper, we do not observe major degradations on linguistic tasks. This suggests that the features learned by our dub-augmented models preserve speech-related information that can be used to, for instance, recognize words or commands. However, the source-separated models' features appear useful for these tasks, which suggests that non-speech features and more general representations of the sound signals may be helpful. We further investigate this below, where our results show that the background noise in one of these datasets (Fluent Speech Commands~\cite{lugosch2019speech}) may provide useful signal for performance.

\paragraph{Is performance on video-only tasks impacted?}
On the visual action recognition tasks, the results from the dub-augmented variants appear similar to the baseline. The baseline performs slightly better on HMDB51 and slightly worse on UCF101. This suggests that the overall video-only performance of the model may not be significantly affected by dub-augmented pretraining, similar to what is shown in Table~\ref{tab:video_performance} for our main model variants.

\begin{table*}[!htb]
\centering
\begingroup\small
\begin{tabular}{lll|lll|lllll}
  \toprule \multicolumn{3}{c}{} & \multicolumn{3}{c}{\textbf{Baselines} (SimCLR)} & \multicolumn{5}{c}{\textbf{Dub-Augmented}} \\
 \cmidrule(l){4-6} \cmidrule(l){7-11}
  & Task & M & \textcolor{black}{AV} & \textcolor{red}{SEP} & \textcolor{brown}{A} & \textbf{ES} & \textbf{FR} & \textbf{JA} & \textbf{EFJ} & \textcolor{brown}{A}\tiny{+\textbf{EFJ}} \\ 
  \midrule
 \midrule
\multirow{3}{*}{\rotatebox[origin=c]{90}{\textbf{SemSp}}} & FlSpComm
\cite{lugosch2019speech} & A & .379 & .400 & \textcolor{gray}{.263} & .391 & \textbf{.410} & .402 & .373 & .368 \\ 
   & SpComm5h
\cite{warden2018speech} & A & .298 & \textbf{.372} & \textcolor{gray}{.144} & .362 & .344 & .325 & .300 & .231 \\ 
   & SpCommFull
\cite{warden2018speech} & A & .471 & .489 & \textcolor{gray}{.162} & .477 & \textbf{.537} & .530 & .491 & .298 \\ 
   \midrule
\multirow{2}{*}{\rotatebox[origin=c]{90}{\textbf{Lang}}} & VoxForge
\cite{maclean2018voxforge} & A & .546 & .516 & \textcolor{gray}{.504} & .580 & .584 & \textbf{.592} & .571 & .543 \\ 
   & VoxLingua10
\cite{valk2021} & A & \textbf{.251\scriptsize{±.045}} & .226\scriptsize{±.033} & \textcolor{gray}{.111\scriptsize{±.012}} & .229\scriptsize{±.016} & .237\scriptsize{±.050} & .246\scriptsize{±.032} & .227\scriptsize{±.043} & .201\scriptsize{±.009} \\ 
   \midrule
\multirow{2}{*}{\rotatebox[origin=c]{90}{\textbf{Act}}} & HMDB51
\cite{kuehne2011hmdb} & V & \textbf{.341} & \textcolor{gray}{.319} & -- & .330 & .324 & .322 & .333 & -- \\ 
   & UCF101
\cite{soomro2012ucf101} & V & .531 & \textcolor{gray}{.496} & -- & .540 & .523 & .538 & \textbf{.542} & -- \\ 
   \bottomrule
   \bottomrule
\end{tabular}
\endgroup
\caption{\textbf{Controlled experiments potential trade-offs: Does dub-augmentation negatively impact performance on linguistic or vision-only tasks?} The tasks in this table include \textbf{Semantic Speech} (FlSpComm~\cite{lugosch2019speech}, SpComm5h~\cite{warden2018speech}, and SpCommFull~\cite{warden2018speech}) and \textbf{Language ID} (VoxForge~\cite{maclean2018voxforge} and VoxLingua10~\cite{valk2021}), and 2 \textbf{Action Recognition} video-only tasks (HMDB51~\cite{kuehne2011hmdb} and UCF101~\cite{soomro2012ucf101}). The results vary and often reflect relatively small differences in either direction, suggesting overall that performance is not majorly affected on language-focused and vision-only tasks.}
\label{tab:tradeoffs_results}
\end{table*}

\section{Examples of Synthetic Counterfactual Pairs}
\label{sec:examples_synthetic}
\begin{figure*}
    \centering
    \includegraphics[width=\textwidth]{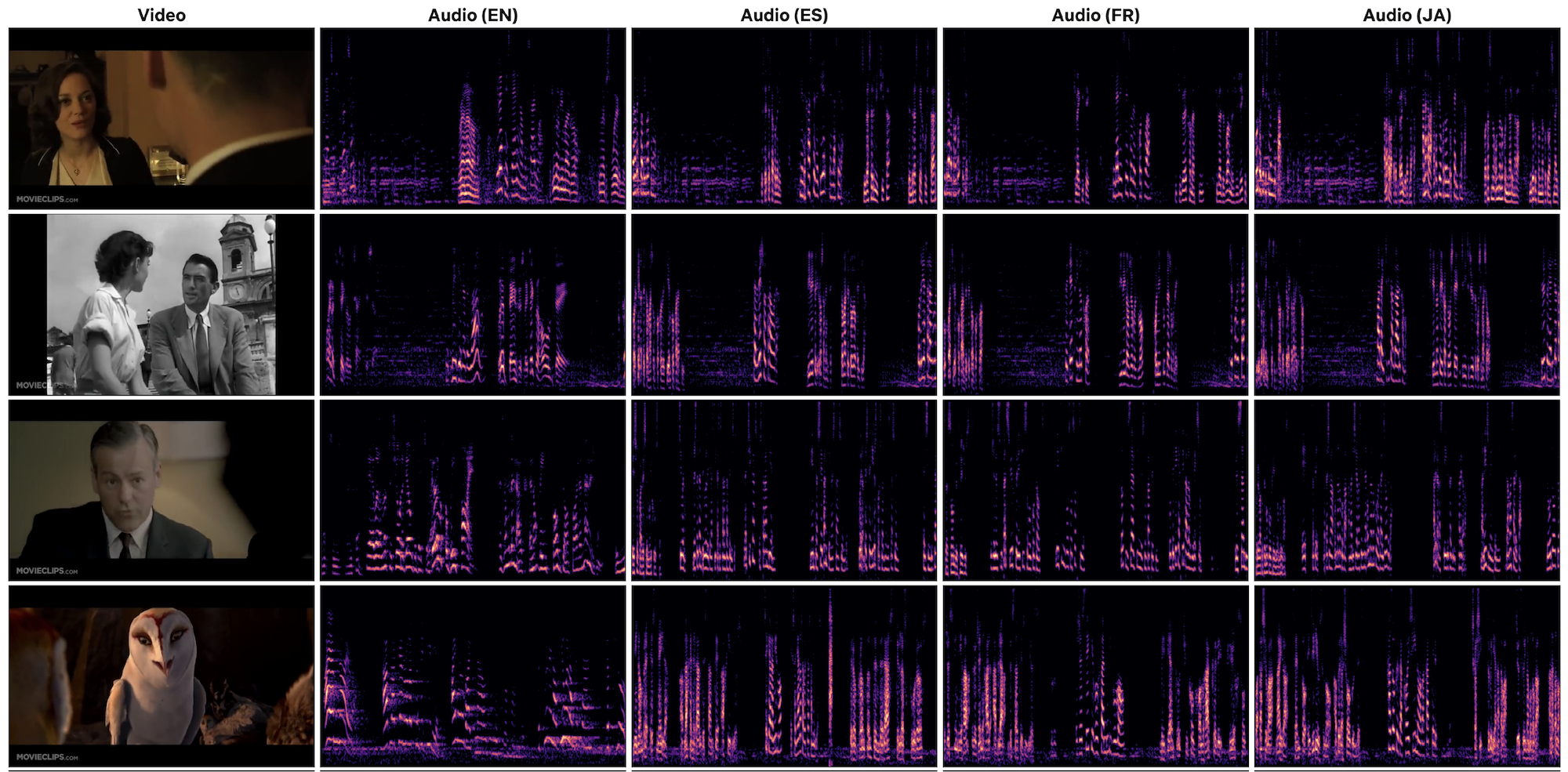}
    \caption{Examples of clips from \textbf{LVU-M}.}
    \label{fig:clips_synthetic}
\end{figure*}

Fig.~\ref{fig:clips_synthetic} highlights clips from a synthetically generated version of the LVU dataset~\cite{lvu2021}, which we refer to as \textbf{LVU-M}, as noted in the main paper. Similar to Fig.~\ref{fig:clips}, the spectrograms show variation and commonalities between alternate audio tracks of the same clip. The examples, arbitrarily selected, show both consistency with the visual (e.g. voices, general timing, etc.) and divergence from it due to artifacts, lack of full acoustic context (e.g. reverberation), and other current limitations of the proposed pipeline. We only show the middle 10 seconds of these clips, to allow easy inspection.

\end{document}